\documentclass[useAMS,usenatbib]{mn2e}
\usepackage{epsfig}

\title[The PKS0347+05 double AGN system]{PKS0347+05: a radio-loud/radio-quiet double AGN system triggered in a major galaxy merger}
\author[C. N. Tadhunter et al.]{C.N. Tadhunter$^{1}$\thanks{E-mail:
c.tadhunter@sheffield.ac.uk}, C. Ramos Almeida$^{2,3}$, R. Morganti$^{5}$, J. Holt$^{5}$, M. Rose$^{1}$, 
\newauthor
D. Dicken$^{6}$, K. Inskip$^{7}$\\
$^{1}$Department of Physics \& Astronomy, University of Sheffield, Sheffield
S3 7RH\\
$^{2}$Instituto de Astrof\'{i}sica de Canarias, C/ V\'ia L\'{a}ctea, E38205 - La Laguna,Tenerife, Spain \\
$^{3}$Departamento de Astrof\'isica, Universidad de La Laguna, E-38205, La Laguna, Tenerife, Spain \\
$^{4}$ASTRON, P.O. Box 2, 7990 AA Dwingeloo, The Netherlands \\
$^{5}$Leiden Observatory, Leiden University, P.O. Box 9513, NL-2300 RA  Leiden, The Netherlands \\
$^{6}$Institute d'Astrophysique Spatiale, Centre universitaire d'Orsay, B\^{a}t 120 -- 121, 91405 ORSAY CEDEX, France \\
$^{7}$Max-Planck-Institut fur Astronomie, Konigstuhl 17, D-69117 Heidelberg,
Germany}
\begin{document}



\maketitle

\label{firstpage}

\begin{abstract}
We present optical, infrared and radio observations of the powerful FRII radio source PKS0347+05 ($z=0.3390$), and demonstrate that it is a rare example of a radio-loud/radio-quiet double AGN system, comprising a weak line radio galaxy (WLRG) separated by 25~kpc (in projection) from a Seyfert 1 nucleus at the same redshift. Our deep Gemini optical images show a highly disturbed morphology, with a warped dust lane crossing through the halo and nuclear regions of the radio galaxy host, tidal tails, and a bridge connecting the radio galaxy to the Seyfert 1 nucleus.
Spectral synthesis modelling of our Gemini optical spectrum of the radio galaxy shows evidence for a reddened young stellar population of age $\le$100~Myr. Further evidence for recent star formation activity in this source is provided by the detection of strong PAH features in mid-IR Spitzer/IRS spectra. Together, these observations support a model in which both AGN have been triggered simultaneously in a major galaxy merger. However, despite the presence of a powerful FRII radio source, and the apparently plentiful supply of fuel provided by the merger, the nucleus of the radio galaxy shows only weak, low ionization emission line activity.
We speculate that the fuel supply to nuclear regions of the radio galaxy has recently switched off (within the last $\sim$10$^6$~yr), but the information about
the resulting decrease in nuclear AGN activity has yet to reach the extended lobes and hotspots of the FRII radio source.  Based on this scenario, we estimate that powerful, intermediate redshift
FRII radio sources have lifetimes of $\tau_{FRII} \sim  5\times10^6$~yr. Overall,
our observations emphasise that the fuelling of AGN activity in major galaxy mergers is likely to be highly intermittent.
\end{abstract}


\section{Introduction}

Understanding the triggering of powerful, radio-loud AGN activity is important
because relativistic jets and lobes of such AGN provide one of the most important forms of AGN-induced feedback: heating the hot gas in the extended X-ray haloes of
the host galaxies and galaxy clusters and preventing it from cooling \citep{mcnamara07}. The jets can also drive shocks and outflows in the warm gas on $\sim$1 -- 10kpc scales \citep{holt08}, directly affecting the star formation histories of the galaxy bulges.

In contrast to the results found for AGN of modest luminosity \citep[e.g.][]{cisternas11,grogin05}, recent deep imaging and spectroscopic observations have provided strong evidence that luminous quasar-like AGN ($L_{bol} > 10^{45}$~erg s$^{-1}$), of both radio-loud and radio-quiet varieties, are triggered in galaxy mergers
\citep{ramos11,ramos12,tadhunter11,villar11,villar12,bessiere12}. However, the triggering of these luminous AGN does not correspond to a single stage of a particular type of merger, with systems observed well before, around and after the time of coalescence of the nuclei of the merging galaxies. In this context, there is clearly an interest in understanding in greater detail the conditions that lead to the triggering of AGN in galaxy mergers.

\begin{table}
\begin{tabular}{lll}
\hline
 &Radio Galaxy &Seyfert 1  \\
\hline
Redshift &$0.33903\pm0.00016$ &$0.33867\pm0.00011$ \\
Spectral Type &WLRG &Sey 1 \\
Radio Morphology &FRII &-- \\
$P_{5GHz}$ (W Hz$^{-1}$) &$4.5\times10^{26}$ &-- \\
$L_{[OIII]}$ (W) &$5.0\times10^{33}$ &$1.6\times10^{34}$ \\
$L_{IR}$ ($L_{\odot}$) &$(2.5^{+0.2}_{-0.6})\times10^{11}$ &--  \\
Stellar mass ($M_{\odot}$) &$1.3\times10^{12}$ &$4\times10^{11}$ \\
\hline
\end{tabular}
\caption{Summary of the general properties of the radio galaxy
PKS0347+05 and companion Seyfert 1 galaxy. Note that the infrared luminosity
$L_{IR}$ was calculated according the the prescription of \citet{sanders96}
and represents the infrared luminosity for the system as a whole (the  far-IR
emission of PKS0347+05 is unresolved and cannot be attributed solely to one
of the two main components). The stellar masses were estimated using the
K-band photometry of \citet{inskip10}.}
\end{table}

In this paper we present a detailed optical, infrared and radio study of the powerful FRII radio source PKS0347+05 (4C+05.16: see Table 1), which is one of the most spectacular merging systems revealed by our recent deep Gemini imaging study of a complete sample of 46 southern 2Jy radio sources \citep{ramos11}. Previous observations of PKS0347+05 produced apparently ambiguous results concerning the
nature of the host galaxy and AGN: while imaging observations showed a diffuse galaxy at the expected position of the radio source host \citep{diserego94}, published optical spectra of PKS0347+05 reveal strong broad line and non-stellar continuum emission characteristic of a broad line radio galaxy (BLRG) or radio-loud quasar \citep{allington91,diserego94} -- apparently inconsistent with the diffuse character of the host galaxy. In section 3.1 we describe new optical imaging and spectroscopy observations which resolve the apparently ambiguous nature of this source, revealing it to be a rare example of a  radio-loud/radio-quiet double AGN. The nature of the AGN activity is investigated in greater detail in section 3.2, while we consider the recent star formation history of the radio source host galaxy in sections 3.3 and 3.4. Finally, in section 4 we discuss the implications of the results for our understanding of the triggering of AGN in galaxy mergers, and present
a new method for determining the lifetimes
of extragalactic radio sources. Throughout this paper we assume a cosmology with $H_0 = 71$~km s$^{-1}$ Mpc$^{-1}$, $\Omega_m = 0.27$, and $\Omega_{\lambda} = 0.73$. This implies
a luminosity distance of 1775~Mpc, and a scale of 4.804~kpc arcsec$^{-1}$, for the redshift of PKS0347+05.

\begin{figure*} 
\epsfig{file=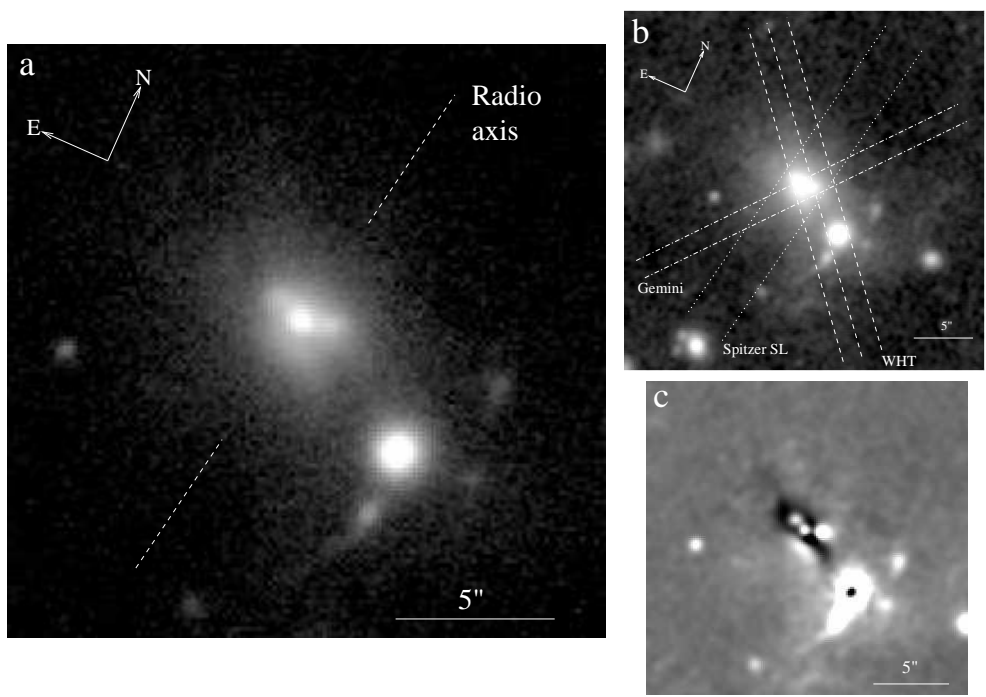, width=13cm}
\caption{Deep imaging observations of PKS0347+05: (a) Gemini r'-band image of PKS0347+05; (b) Gemini r'-band image at higher contrast with
overlays showing the positions of the slits used for the WHT, Gemini
and Spitzer/IRS spectroscopic observations. See \citet{ramos11} for details
of the reduction of the optical imaging data; (c) Gemini r'-band following subtraction of Sersic law model fitted to the radio galaxy component.}
\end{figure*}

\section{Observations}

New optical long-slit spectroscopic observations for PKS0347+05 were taken in February 2007 using the ISIS spectrograph on the 4.2m William Herschel Telscope (WHT) on La Palma, and in December 2008 using the GMOS instrument on the Gemini South telescope on Cerro Pach\'on, Chile. In both cases the slit was aligned along the parallactic angle, in order to avoid wavelength-dependent slit losses due to differential atmosperic refraction. Details of the spectroscopic observations are given in Table 2, while the slit positions of the various
spectroscopic observations are shown overlayed on our Gemini-S r'-band image in 
Figure 1b.

\begin{table*}
\begin{tabular}{lllllllll}
\hline
Telescope &Instrument &Grating &Date &Exposure &PA &Slit Width &Seeeing &$\Delta\lambda$\\
& & & &(s) &(degrees) &(") &FWHM(") &(\AA) \\ 
\hline
WHT &ISIS &R300B &9/02/07 &900 \& 2700 &40 &1.5 &1.0 &5.5 \\
& &R158R &9/02/07 &900 \& 2700 &40 &1.5 &1.0 &10.5 \\
Gemini-S &GMOS &B600 &23/12/08 &3600 &140 &1.5 &0.57 &7.0  \\
&  &R400 &23/12/08 &1800  &140 &1.5 &0.57 &13.7 \\
Spitzer &IRS &SL &8/10/08 &480 &169 &3.6 &-- &260\\
& &LL &8/10/08 &480 &85 &10.5 &-- &100\\
\hline
\end{tabular}
\caption{Details of the long-slit spectroscopic observations of PKS0347+05. The final
column gives the spectral resolution. In the case of the WHT observations
the shorter and longer exposure times correspond to observations taken with
the slit centred on the eastern and western sources respectively.}
\label{tab:list}
\end{table*}

In the case of the WHT data, observations were taken simultaneously with both the red and blue arms of the ISIS dual arm spectrograph, using the R300B and R158R gratings with a 1.5 arcsecond slit. Initially the slit was centred on the diffuse galaxy along the axis of the large-scale FRII radio source. However, the spectrum of the this galaxy showed only weak narrow emission lines, with no sign of the strong broad lines reported in previous studies of this source \citep{allington91,diserego94}. Therefore a spectrum was also taken with
the slit centred on the object $\sim$5 arsceconds to the south west of the nucleus of the diffuse galaxy, because this has the compact mophology that would generally be expected for a broad line AGN. Indeed, the spectrum of the compact object proved to have strong broad lines similar to those detected in the previous observations of PKS0347+05. 

Motivated by the spectacular optical morphology of PKS0347+05 revealed
by our deep Gemini-S r'-band image \citep[][see Figure 1]{ramos11}, a deep spectrum of the diffuse galaxy along the radio axis was taken using Gemini-S/GMOS in spectroscopic mode, in order to improve on the S/N of the WHT/ISIS data and allow spectral synthesis modelling of the stellar populations. The Gemini-S/GMOS observations were made using the B600 and R400 gratings, with the 1.5 slit centred on the nucleus
of the diffuse galaxy. Although the B600 and R400 observations were not taken simultaneously, the observations with the two gratings were interleaved, in order to ensure uniform observing conditions and similar airmass and parallactic angle for the red and blue spectra.  

The reduction of the WHT data followed the standard steps of bias subtraction, flat fielding, wavelength calibration, flux calibration, and 
correction for the tilt in the long-slit spectrum in the spatial direction; all the reduction was done using the NOAO package in IRAF\footnote{IRAF is distributed by the National Optical Astronomy Observatory, which is
operated by the Association of Universities for the Research in Astronomy, Inc., under
cooperative agreement with the National Science
Foundation (http://iraf.noao.edu/).}. The reduction of the Gemini data was similar, except that the IRAF Gemini package was used to mosaic the three GMOS CCD images and subtract the bias before  proceeding with the wavelength and flux calibration. All the spectra were corrected for Galactic reddening \citep[$E(B-V)=0.274$ from][]{schlegel98} prior to further analysis. 

Based on the measurement of night sky lines recorded in the long-slit spectra, the accuracy of the wavelength calibration is better than 0.5\AA\ for both the WHT and the Gemini spectra, while comparisons between observations of several flux standard stars demonstrates that the relative flux calibration is accurate to within $\pm$5\%. Reassuringly, no scaling was required to match the flux levels of the red and blue arm WHT/ISIS spectra, and the B600 and R400 Gemini/GMOS-S spectra, in the wavelength regions where they overlap; the flux levels of the blue and red spectra agree to within 5\%.

Note that the reduction of r'-band Gemini images and Spitzer/IRS spectra discussed
in the following sections are described in detail in \citet{ramos11} and \citet{dicken12} respectively.

\section{Results}
\subsection{Optical morphology}

Figure 1 shows our deep r'-band image of the PKS0347+05 system, clearly
revealing that it has a complex morphology comprising two galaxies with projected
separation 5.3 arcsec (25~kpc), a diffuse and irregular low surface brightness
envelope that extends 15 arcsec (70~kpc) to the north west of the eastern galaxy, and a system of knot and tail features centred on the western galaxy
that are co-aligned in the north-south direction.

To examine the near-nuclear structure in the eastern galaxy in more detail, we
have subtracted a model for its surface brightness profile. We performed the modelling using GALFIT version 3.0.2 \citep{peng02,peng10}\footnote{GALFIT is a well-documented two-dimensional fitting algorithm which allows the user to simultaneously 
fit a galaxy image with an arbitrary number of different model components, in order to extract structural 
parameters of the galaxy. The model galaxy is convolved with a point spread function (PSF) and, using the 
downhill-gradient Levenberg-Marquardt algorithm, is matched to the observational data via the minimization 
of the $\chi^2$ statistic.}. First we derived a PSF profile by extracting
the  2D images of several stars in the GMOS-S image, normalizing to 
unit flux and taking an average profile weighted by the signal-to-noise ratio of the component 
extracted stellar profiles. The system was then modelled over a 70$\times$70 arcsec$^2$ 
area, using a S\'ersic profile (S\'ersic 1963) to fit the eastern galaxy, and a point source (PSF profile) to fit the compact western galaxy. The best-fitting S\'ersic profile index for the eastern galaxy
was $n=2.8$.  All model parameters, including the centroids of the two galaxies, were 
allowed to vary freely. We also left the residual background level as an additional free parameter.

Figure 1c shows the result of the GALFIT model subtraction. Clearly the near-nuclear structure of the eastern galaxy is highly complex, with
warped dust lane that
threads through the nuclear regions and eastern halo of the galaxy, and has an extent of 2.3 arc sec (11 kpc) to the west, and 5.1 arc sec (25 kpc)
to the east. It is notable that
the western extension of the dust lane 
points towards the nucleus of the western galaxy. There is also evidence 
for a second dust lane cutting through the extended halo of the system $\sim$5
arc sec to the north of the western nucleus.

The nucleus of the western galaxy is only marginally resolved
(FWHM$\sim$0.6 arcsec) at the resolution of our Gemini data. On the other hand, the nuclear region of the eastern galaxy is resolved into a central compact component, with high surface brightness condensations 1.1 arc sec (5.3~kpc) to the
east in the direction of the dust lane, and 1.2 (5.8~kpc) arc sec to the north west (see Figure 1c). The fact that the
latter feature is also visible in the K-band image of \citet{inskip10}, suggests that it may represent a third nucleus in the system. There is also a region of more diffuse emission that runs to the south of, and parallel to, the dust lane. It is possible that all of these brighter features
represent regions of star formation that are associated with, and partially obscured by, the dust lane.

We note that \citet{inskip10} find that a substantial contribution from a point source (20\%) is required in order to fit the K-band light profile of the western galaxy adequately, but that it is possible to fit the light profile of the eastern galaxy using a Sersic law (index= 4) without any point source component. This analysis also shows that the ratio of the total K-band flux of the eastern to the western galaxy is $\sim$3:1.

Overall, the PKS0347+05 system shows a high degree of morphological disturbance at optical wavelengths, consistent with the idea that it represents a major galaxy merger observed in the pre-coalescence stage, before the nuclei of the precursor galaxies finally coalesce. 

\subsection{The nature of the AGN activity in PKS0347+05}

As described in the introduction, the AGN classification of PKS0347+05 has been ambiguous in the past: while the spectroscopic studies classify it as a broad line radio galaxy (BLRG) on the basis of the detection of a strong broad H$\alpha$ line, this is apparently at odds with the diffuse nature of the host galaxy.

\begin{figure} 
\epsfig{file=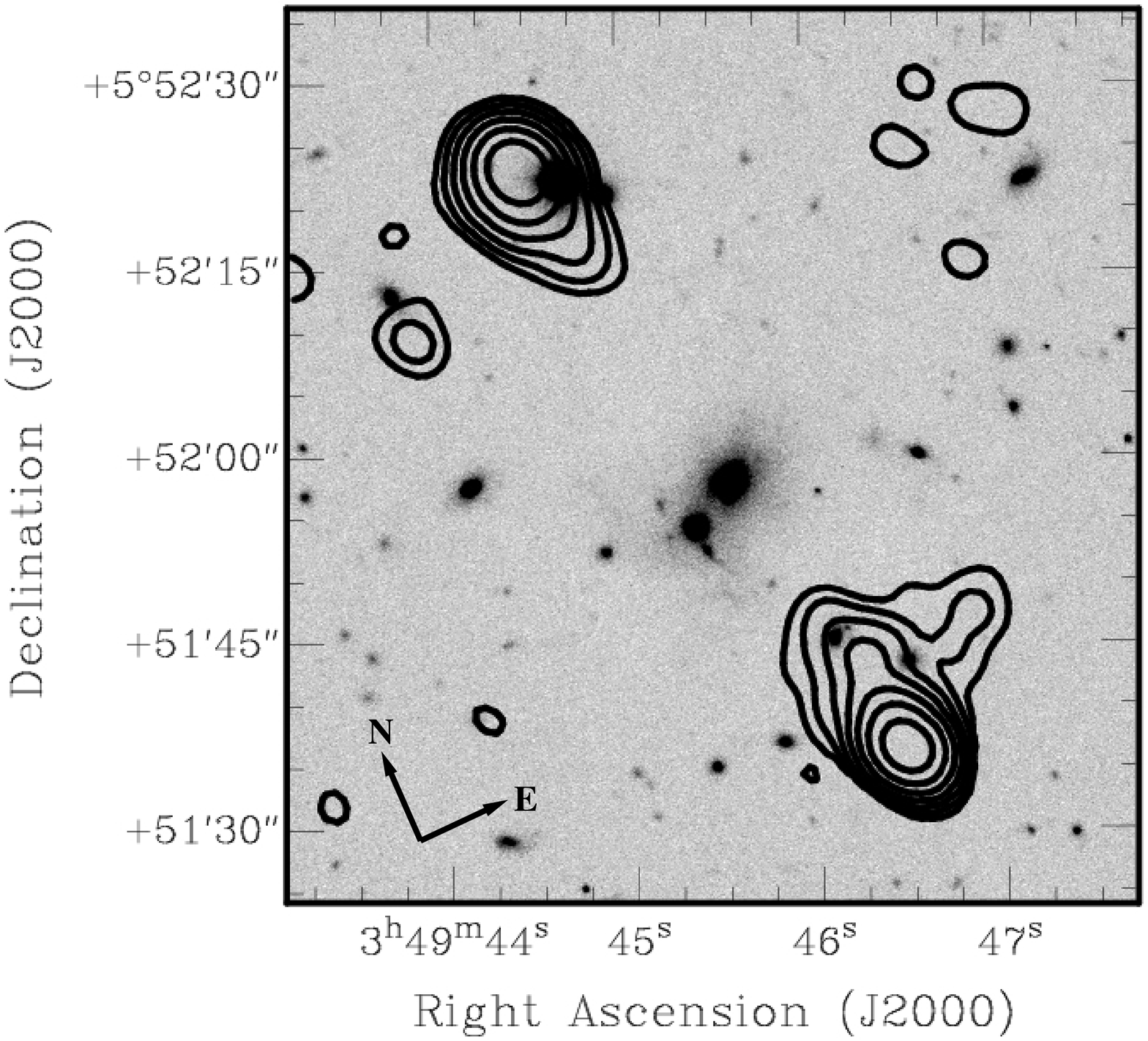, width=8.5cm}
\caption{Gemini r'-image image of PKS0347+05 (greyscale) overlayed with the 5GHz radio map of \citet{morganti93} (contours). }
\end{figure}

Figure 2 shows an overlay of the 5~GHz VLA radio map  on our r'-band Gemini image. The 5~GHz
map reveals a classical FRII morphology aligned along PA170, with strong, relatively undistorted, edge-brightened radio lobes. The total diameter of the radio source
is 326$\pm$24 kpc. Although no radio core is detected at either 5~GHz or 22~GHz \citep{morganti93,dicken09}, the fact that the line connecting the
highest surface brightness regions in the two radio lobes falls with 0.5 arcsec
of the
nucleus of the eastern galaxy provides strong evidence that the radio jets
originate in the eastern galaxy \citep[see also][]{diserego94}. While
we cannot entirely rule out the idea that radio jets originate in the 
western nucleus, this is highly unlikely given the undistorted nature of the radio structure and the close positional agreement between the radio axis and the nucleus of the eastern galaxy.

The apparently ambiguous nature of the optical AGN classification is resolved by our new spectroscopy. In the course of our 2007 WHT observations we made spectroscopic observations of both the diffuse eastern galaxy and its more compact companion to the west (see Figure 3). The eastern galaxy shows weak, narrow H$\alpha$ and [NII] emission lines, and [OIII]$\lambda$5007 emission that is either
weak or absent. In contrast, our spectrum of the western nucleus reveals broad H$\alpha$, H$\beta$ and FeII emission lines that are characteristic of quasars or Seyfert 1 nuclei, although its relatively low [OIII] luminosity (see Table 3) is more typical of Seyfert 1 nuclei than quasars \citep[see][]{zakamska03}. It is also notable that the general character of the spectrum of western nucleus, with its low equivalent width [OIII] emission and relatively strong broad FeII blends, has more in common with radio-quiet that radio-loud AGN. Indeed, such spectral characteristics are rare in steep radio spectrum, radio-loud AGN \citep{boroson92}.

\begin{figure} 
\epsfig{file=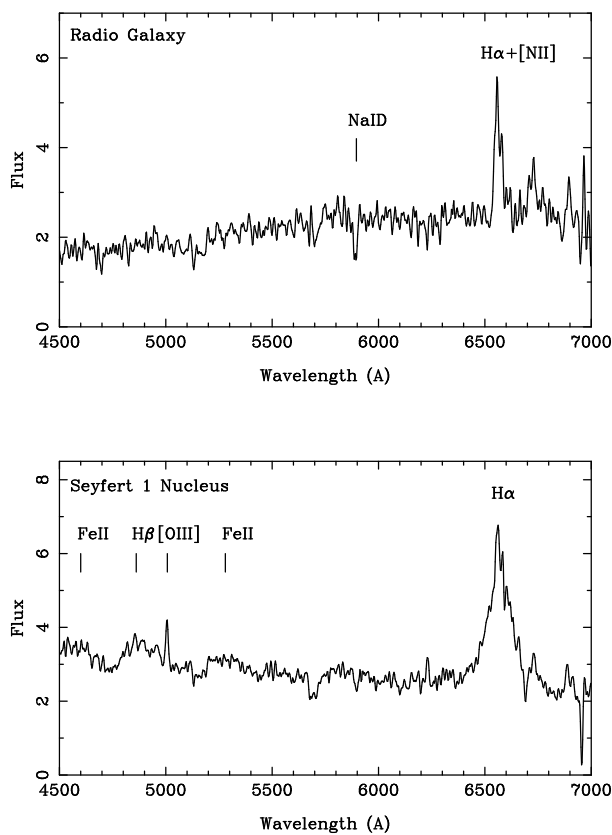,width=8cm}
\caption{WHT spectroscopic observations of the PKS0347+05 system. Top: spectrum of the nucleus of the diffuse galaxy to the east (1.5$\times$2.2 arsec extraction aperture). Bottom: spectrum of the compact object to the west
(1.5$\times$1.6 arsec extraction aperture). The fluxes are
in units of $10^{-17}$~erg cm$^{-2}$ s$^{-1}$ \AA$^{-1}$. }
\end{figure}

{\it We contend that previous spectroscopic studies of PKS0347+05 have reported observations of the nucleus of the western companion galaxy (this
would appear brighter in target acquisition systems), rather than the true
(diffuse) radio source host galaxy to the east.} This explains the previous classification of this
source as a BLRG. In the following we refer to the eastern galaxy as the radio galaxy, and the western galaxy as the Seyfert 1 nucleus. 

Further clues to the nature of the AGN in the radio galaxy host are provided by our deeper Gemini/GMOS spectrum of the nucleus of the eastern galaxy (Figure 4). This reveals weak [OIII]$\lambda\lambda$5007,4959, H$\beta$ and [OII]$\lambda$3727 emission lines as well as a strong stellar continuum. Given the low equivalent width of its [OIII]$\lambda$5007 emission line ($\le$10\AA\,) this object qualifies as a weak line radio galaxy (WLRG) according to the criterion of \citet{tadhunter98}. Table 3 summarises the line ratio information for the nucleus of the eastern galaxy. The generally low ionization character of the nuclear spectrum is characteristic of WLRG in general. However, this object falls in the ``composite'' region of the [OIII]/H$\beta$ vs 
[NII]/H$\alpha$ diagnostic diagram shown in Figure 5, suggesting a significant contribution from stellar, rather than AGN, photoionization \citep{kewley06}. Therefore the strength of any AGN-photoionized emission line component in the eastern nucleus is likely to be even weaker than estimated on the basis of the total [OIII] emission line luminosity. It is notable that, despite the possible contribution from stellar photoionization, the [OIII] luminosity of PKS0347+05 is two orders of magnitude lower than radio galaxies of similar redshift and radio power in the 2Jy sample described in \citet{tadhunter98}; it is by far the highest redshift WLRG in the 2Jy sample --- the next highest redshift 2Jy WLRG is PKS1648+05 (Her A) at $z=0.158$.

\begin{table}
\begin{tabular}{lll}
\hline
Line ratio &Value &Instrument\\
\hline
$[OII](3727)/[OIII](5007)$ &1.60$\pm$0.8 &Gemini/GMOS \\
$H\delta/H\beta$ &0.24$\pm$0.06 &Gemini/GMOS \\
$H\gamma/H\beta$ &0.48$\pm$0.07 &Gemini/GMOS \\
$[OIII](5007)/H\beta$ &1.04$\pm$0.09 &Gemini/GMOS \\
$[NII](6584)/H\alpha$ &0.69$\pm$0.12 &WHT/ISIS \\
&0.75$\pm$0.14 &Gemini/GMOS \\
$[NeIII](15.6)/[NeII](12.8)$ &0.39$\pm$0.13 &Spitzer/IRS \\
\hline
\end{tabular}
\caption{Diagnostic emission line ratios for the nuclear regions
of the PKS0347+05 radio galaxy, measured from the Gemini, WHT and
Spitzer spectra. Note that, in the case of the Gemini spectra,
the best-fitting stellar continuum model was subtracted from
the spectra prior to measuring the line ratios. Further details
of the Spitzer spectra are given in \citep{dicken12}. Given
that the longer wavelengths ($>$6000\AA, in the rest frame) of the
Gemini spectrum
are likely to be affected by fringing, we also present the
$[NII](6584)/H\alpha$ ratio measured from the WHT spectrum. }
\end{table}

\begin{figure} 
\epsfig{file=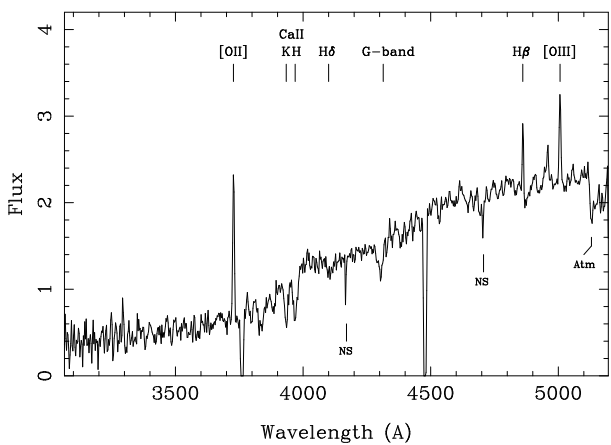,angle=-90,
width=8cm}
\caption{Gemini/GMOS spectrum of the PKS0347+05 radio
galaxy (1.5$\times$1.45 extraction aperture). Note that the dips in the spectra
at $\sim$3760\AA\, and $\sim$4480\AA\, are due to the gaps between the 
CCD detectors used in GMOS instrument. The spectrum has been corrected for
Galactic reddening. The fluxes are
in units of $10^{-17}$~erg cm$^{-2}$ s$^{-1}$ \AA$^{-1}$.}
\end{figure}

\begin{figure} 
\epsfig{file=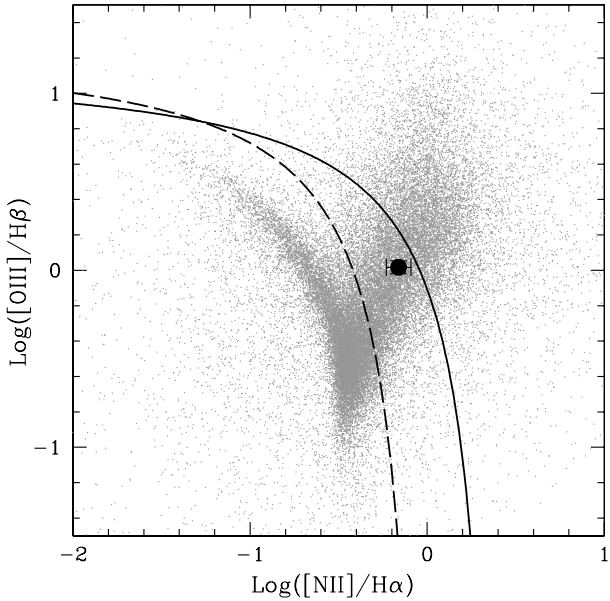,width=8cm}
\caption{$[OIII]/H\beta$ vs $[NII]/H\alpha$ diagnostic plot showing the
position of the nucleus of PKS0347+05 (large black circle) relative to
SDSS detected galaxies (grey dots) from SDR8.  Points below the dashed line
represent HII regions galaxies dominated by stellar photoionization, while
points above the black solid line represent objects dominated by AGN photoionization \citep{kewley06}. Note that
PKS0347+05 falls in the composite zone between these two lines, indicating a 
mixture of AGN and stellar photoionization.}
\end{figure}

\begin{figure}
\epsfig{file=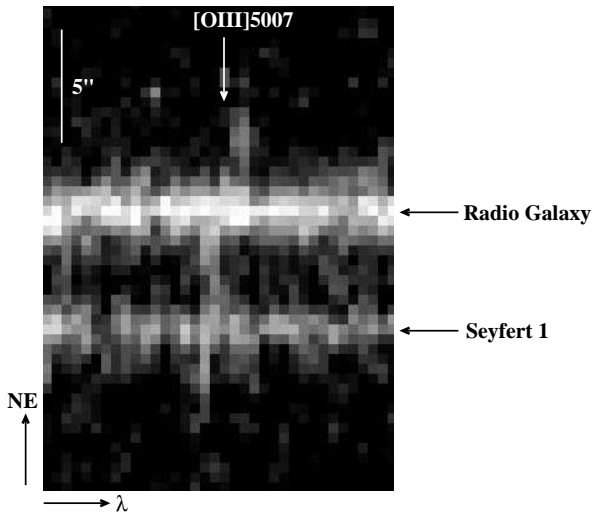,width=8cm}
\caption{Cut out of the PA40 WHT long-slit spectrum, showing the faint, extended [OIII]$\lambda$5007 emission that is detected out to a maximum radius of 9.2 arcseconds of the SW and 4.4 arcsec to the NE of the radio galaxy nucleus. Although the slit is centred on the radio galaxy, it encompasses some emission
from the western galaxy --- the slit passes $\sim$1.5 south of the Seyfert 1
nucleus (see Figure 1b).}
\end{figure}

Although the [OIII] emission is not detected directly in the nuclear regions of the radio galaxy in our low S/N WHT spectrum, we do detect weak extended [OIII] emission in the WHT spectrum along PA40 that extends 9.2 arcsec (44~kpc) to the SW and 4.4 arcsec (21~kpc) to the NE of the nucleus (see Figure 6). The velocity pattern along this PA has the characteristics of a rotation curve, with the 
extended [OIII] emission on the SW side of the nucleus blueshifted by 10.9$\pm$0.8\AA\, (8.1$\pm$0.6\AA, or 485$\pm$36~km s$^{-1}$ in the radio galaxy rest frame) relative to that on the NE side; this velocity pattern is likely to be associated with the rotation of the disk-like structure/dust lane visible in our optical r'-band image. The average of the wavelengths of the [OIII] emission on either side of the nucleus provides an estimate of the systemic redshift of the radio galaxy of $z_{rg}=0.33903\pm0.00016$. This agrees within the measurement uncertainties with
the redshift estimated from a Gaussian fit to the [OIII] emission
detected in our nuclear Gemini spectrum (Figure 4: $z_{rg}=0.33895\pm0.00008$). For comparison, the redshift of the Sy1 nucleus implied by the centroid wavelength of a Gaussian fit to its [OIII] emission is 
$z_{sy1}=0.33867\pm0.00011$. Thus the  rest frame radial velocity difference between radio galaxy and the Sy1 nucleus is small: $\Delta V = -80\pm44$~km s$^{-1}$.

\subsection{The mid- to far-IR characteristics}

In Figures 7a and b we show the mid-IR 24$\mu$m and 70$\mu$m monochromatic luminosities plotted against [OIII] emission line luminosity for the 3CRR and 2Jy samples described in \citep{dicken10}, with the position of PKS0347+05 highlighted. It is clear that PKS0347+05 falls well above the observed correlations at both wavelengths, implying a substantial contribution from a starburst heated dust component \citep{dicken09,dicken10}. Further evidence for dust obscured star formation in PKS0347+05 is provided by its relatively red mid- to far-IR (MFIR) colour (F(70)/F(24)$=$8.8$\pm$1.3), which is characteristic of starburst objects, but is much redder than measured in AGN-dominated radio galaxies (Dicken et al. 2009, 2010). Using the Spitzer photometric data presented in \citet{dicken09} we estimate that the mid- to far-IR (MFIR) luminosity of the PKS0347+05 system as a whole is $L_{IR} = (2.5^{+0.2}_{-0.6})\times10^{11}$~L$_{\odot}$ -- implying that it is a luminous rather than an ultraluminous infrared galaxy \citep{sanders96}.

\begin{figure} 
\epsfig{file=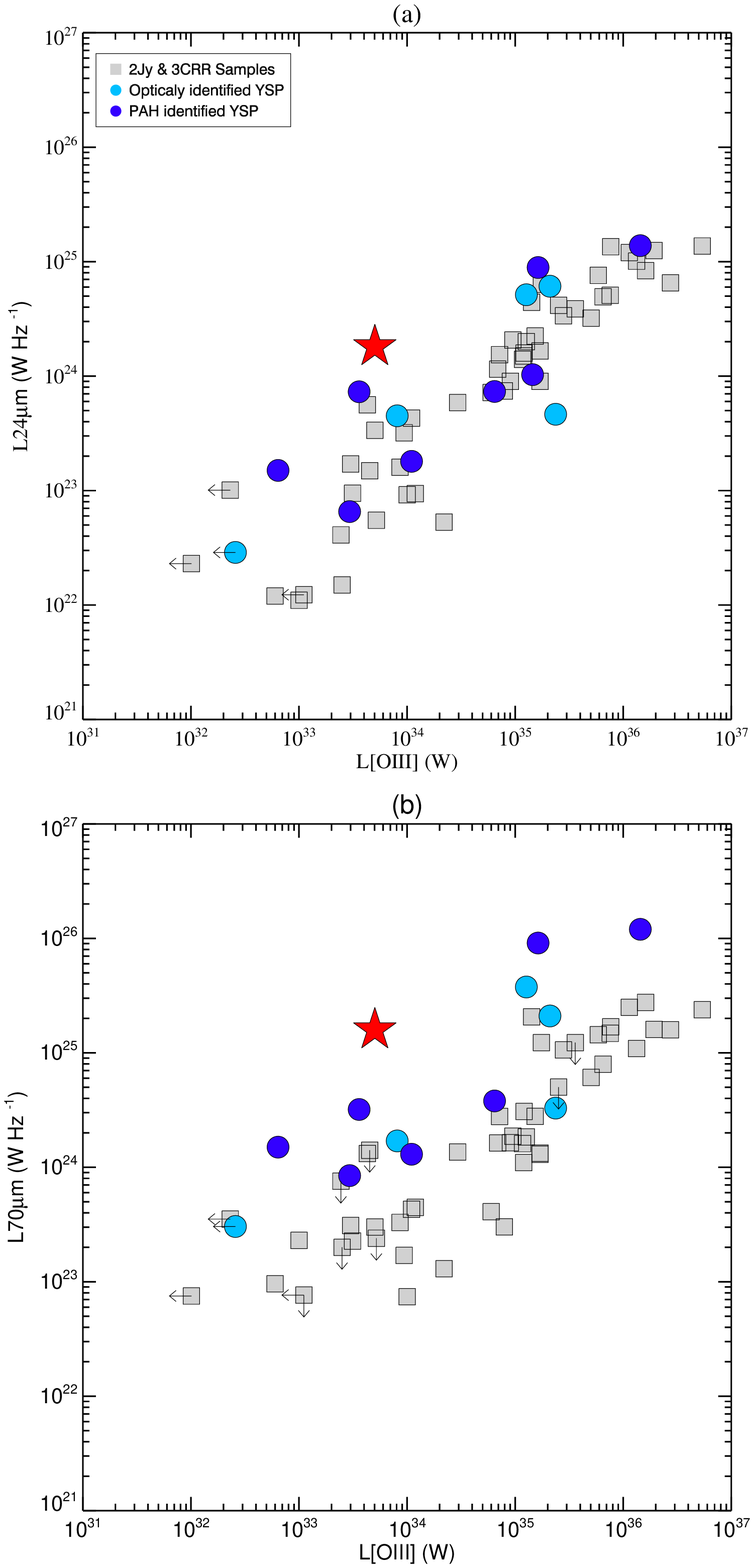,width=8cm}
\caption{Position of PKS0347+05 (red star) on the correlations between 
$L_{24\mu m}$ vs $L_{[OIII]}$ (top) and $L_{70\mu m}$ vs $L_{[OIII]}$ (bottom)
for samples of 2Jy and 3CRR radio sources \citep[see][for details]{dicken09,dicken10},
with objects showing optical and PAH evidence for recent star formation
identified with circles. Note that the PKS0347+05 system has mid- to far-IR (MFIR) luminosities
that are an order of magnitude higher than typical radio galaxies in
the 3CRR and 2Jy samples of similar [OIII] luminosity.}
\end{figure}

The overall impression of prodigious star formation activity, but weak AGN activity, is reinforced by our Spitzer IRS spectrum of the radio galaxy (Figure 8, see Dicken et al. 2012 for details), which shows strong PAH 6.6, 7.7, 11.3$\mu$m emission bands and a moderately strong [NeII]$\lambda$12.8$\mu$m line, but no sign of lines such as [NeV]$\lambda$14.3$\mu$m or [OIV]$\lambda$25.9$\mu$m which are characteristic of AGN. Moreover, the [NeIII]$\lambda$15.6$\mu$m line is barely detected. We further note that the luminosity of the 11.3$\mu$m PAH feature
detected in PKS0347+05 ($L_{11.3} \sim 3\times10^{37}$W) is the second highest in the 2Jy sample \citep{dicken12}. Strong PAH features are rare in powerful radio galaxies; they are only detected in 30\% of the objects in the 3CRR and 2Jy samples studied by \citet{dicken12}.

\subsection{The star formation history and stellar mass of the host galaxy}

The optical emission line spectrum, MFIR excess, red MFIR colours, and strong PAH  features all provide evidence for recent star formation activity in PKS0347+05. We can gain further insights into the star formation history of the
radio galaxy by modelling the deep Gemini spectrum of its nucleus (Figure 4). This spectrum has a composite character, showing absorption features that are characteristic of relatively old stellar populations (e.g. G-band, CaII K), as well a weak H$\delta$ Balmer line absorption and a UV excess that are characteristic of much younger stellar populations.

For this purposes of modelling the spectrum we have used CONFIT, which is a purpose written IDL based code \citep[see][]{robinson00,tadhunter05}. Employing a minimum $\chi^2$ technique to find the best fit to spectra, CONFIT attempts to model each spectrum with the minimum number of components. In the case of PKS0347+05 we used two components: an old stellar population (OSP) with an age 12.5 Gyr, and instantaneous burst young stellar population with age in the range 0.001-5.0 Gyr and reddening in the range $0.0<E(B-V)<1.6$
(solar metallicity model templates taken from Bruzual \& Charlot 2003). {We assumed a simple screen geometry for the reddening
dust and used a \citet{calzetti00} extinction law for all the stellar populations.}

\begin{figure*} 
\epsfig{file=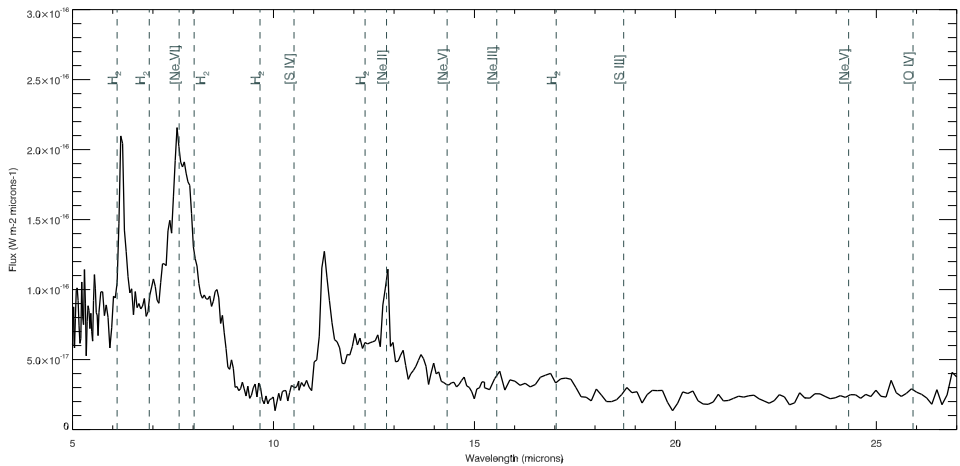,width=13cm}
\caption{Spitzer IRS spectrum of PKS0347+05 with the positions of various fine-structure and H$_2$
emission lines identified. Note the strong PAH emission band features at 6.6, 7.7, 11.3$\mu$m. Details of the reduction of the IRS data are presented in \citet{dicken12}.}
\end{figure*}

Prior to the modelling it was important to gauge the level of any nebular continuum component, since this component is known to make a significant contribution
to near-UV continua of narrow line AGN with high equivalent width emission lines \citep{dickson95,tadhunter02}. In the case of PKS0347+05 we find that, based on the
H$\beta$ flux corrected for underlying stellar absorption, the nebular continuum is negligable, contributing $<$5\% of the continuum below 3500\AA. Moreover, given that the level of AGN activity indicated by the narrow lines is relatively low in this object, it is unlikely  that AGN-related continuum components such as scattered or direct AGN emission contribute
significantly to the optical/UV \citep[see][]{tadhunter02}. Therefore no correction for nebular continuum was deemed necessary, and we did not include components in our model to specifically take into account a scattered or direct AGN continuum component.

In carrying out the modelling, 52 continuum bins of 30{\AA} width were used. These were chosen to avoid emission lines, telluric absorption features, gaps between the GMOS CCD detectors, and image defects. The wavelength range for the modelling was restricted to the rest wavelength range $\sim3000-6000${\AA}, in order to avoid the red end of the Gemini/GMOS spectrum which is strongly affected by fringing. The spectral bin $4050-4080${\AA} was
selected for normalisation. CONFIT works by scaling the flux from the model components so that the total flux incorporated in the model is always less that 125\% of the observed flux in the normalising bin. The code then calculates the minimum $\chi^2$ for each combination of components using different relative fluxes between the components. 
A relative flux calibration error of $\pm5\%$ was assumed for modelling the continuum of PKS0347+05. 

Initially we assumed an unreddened OSP. This assumption is justified on the basis that old stellar populations in the foreground of the halo of the galaxy are unlikely to suffer major extinction by the near-nuclear dust lane, while old stellar populations in the stellar halo that are behind the dust lane along the line of sight are likely to be substantially extinguished and not contribute significatly to the total light of the OSP component.
We find that, regardless of the degree of reddening in the OSP component, the inclusion of a YSP is essential for modelling the optical/UV continuum of PKS0347+05. The contours of $\chi^2_{red}$ for the unreddened OSP models show three minima, with three combinations of YSP age and reddening providing a viable fit to the overall shape of the spectrum ($\chi^2_{red} < 1$): 0.005 -- 0.01~Gyr with $1.1 < E(B-V) < 1.4$; 0.06 -- 0.1~Gyr with $1.1 < E(B-V) < 1.4$; and 0.8 -- 1.2~Gyr with $0.2 < E(B-V) < 0.4$. Detailed inspection of the fits to the stellar absorption features rules out the solutions with older age YSP (0.8 -- 1.2~Gyr) because they substantially over-predict the CaII~K absorption line strength. However, we cannot distinguish between the two younger age YSP models based on the detailed fits. A major problem with the remaining viable models is that they require highly reddened YSP components, implying total YSP masses and star formation rates of the YSP components that are implausibly high ($M_{ysp}>10^{11}$~M$_{\odot}$, and $SFR > 10^3$~M$_{\odot}$ yr$^{-1}$).

\begin{figure} 
\epsfig{file=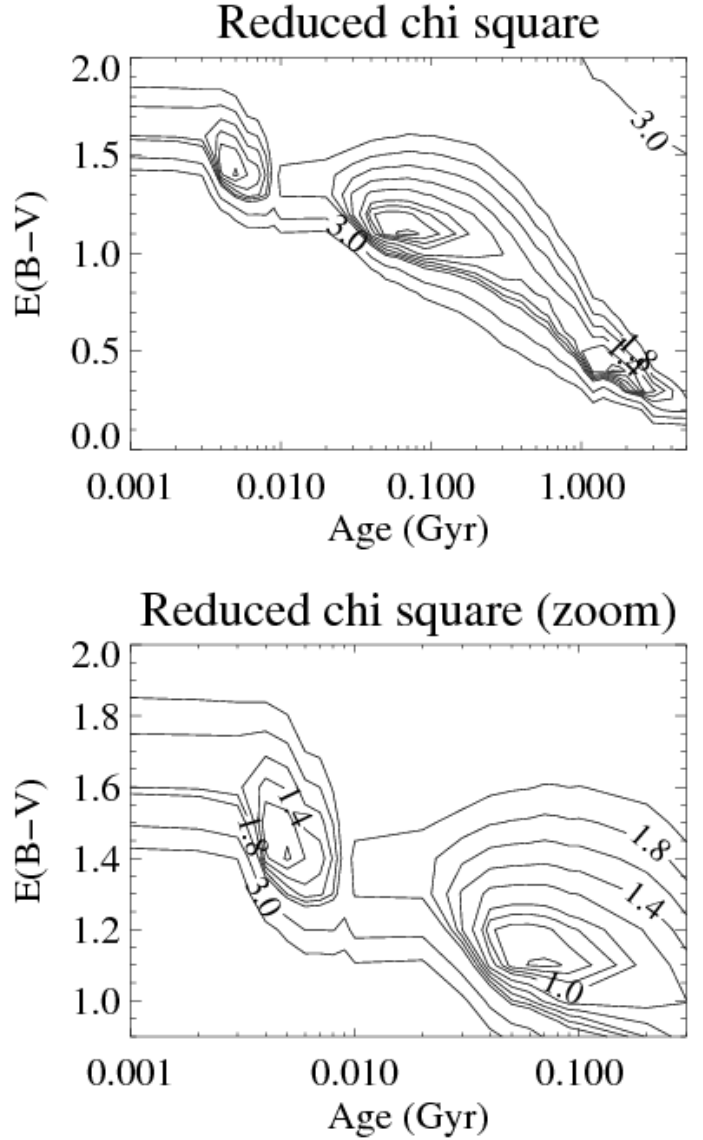,width=8cm}
\caption{Spectral synthesis modelling of the nuclear spectrum of
PKS0347+05, showing the contours of minimum $\chi^2$ for different
combinations of the age and reddening (E(B-V)) of the YSP. The upper
plot shows the full range of age and reddening used in the modelling, while
the lower plot shows an expanded view of the younger stellar ages.}
\end{figure}

Given the problems with the zero reddening OSP models, we then experimented with models in which the OSP are lightly reddened by $E(B-V) = 0.15$ and $E(B-V) = 0.3$. In the case of the $E(B-V) = 0.3$ OSP models we  obtain viable fits to both the overall SED and detailed absorption features for YSP ages $t_{ysp} < 0.1$~Gyr and YSP reddening $0.4 < E(B-V) < 1.0$\footnote{Note that we assume that the YSP is more highly reddened than the OSP (i.e. $E(B-V) > 0.3$).}, while the viable models with OSP reddened by  $E(B-V) = 0.15$ have similar YSP ages and  $0.8 < E(B-V) < 1.1$ (see Figures 9 and 10 for examples of the
model fits). Although they produce similar estimates of YSP ages, the major advantage of these lightly reddened OSP models is that, because the YSP reddening is lower, and the YSP contribute less to the flux in the normalising bin, the masses of the YSP and the star formation rates are more plausible ($4\times10^9 < M_{ysp} < 5\times10^{10}$~M$_{\odot}$, $40 < SFR < 500$). On the basis of these models we estimate that the YSP contribute between 0.6 and 10\% (depending on the precise model) of the total stellar mass in the central regions of the galaxy covered by our spectroscopic aperture.  

\begin{figure*} 
\epsfig{file=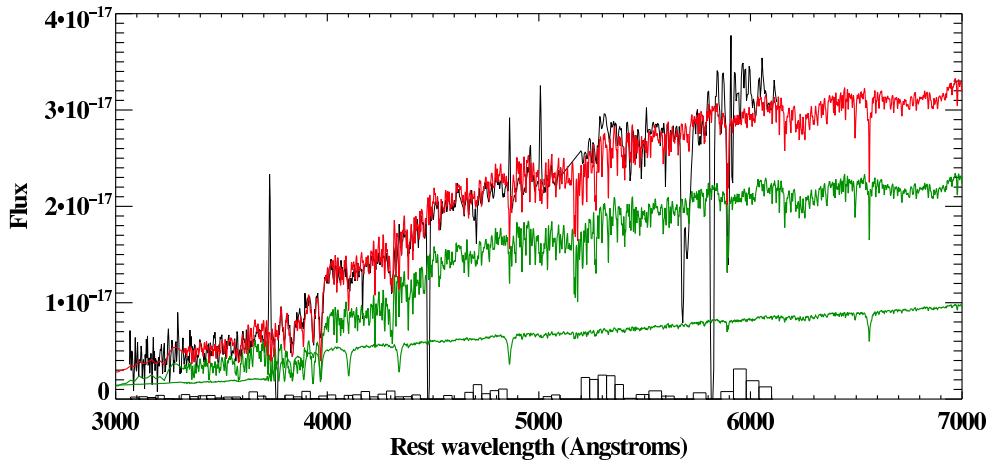,angle=0,width=13cm}
\caption{Example of a model fit to the nuclear spectrum of
PKS0347+05. The model comprises a combination of the lightly reddened
OSP ($E(B-V) = 0.15$: upper green curve) and a 0.08~Gyr YSP with a 
higher reddening ($E(B-V) = 1.0$: lower green curve). The black line
shows the observations, while the red line shows the model obtained 
by combining 
the OSP and YSP components; the histogram boxes at the bottom of the plot show
the differences between the fluxes of the model and the data in each of the 
wavelength bins used for the modelling. }
\end{figure*}

It would be difficult to be more precise about the ages of
the YSP in PKS0347+05, because of the uncertainties in the
templates used in the modelling, and also because the starburst may not
in fact be instantaneous. However, the estimated YSP ages 
($0.05 < t_{YSP} < 0.1$) overlap with the range of
typical lifetimes of extragalactic radio sources, as estimated
on the basis of spectral aging and dynamical arguments \citep[e.g.][]{alexander87,parma99}, as well as the rate of occurrence
of WLRG in the radio source population (see section 4.2 below). 
Therefore it is plausible
that the AGN and jets in PKS0347+05 were triggered during
the most recent phase of merger-induced starburst activity. Certainly, there
is no evidence for a substantial delay in the triggering of the AGN activity, as found
in some studies of lower luminosity radio galaxies \citet{tadhunter05,tadhunter11} and radio-quiet
AGN \citep{wild10}.

We have also used our spectral synthesis modelling results in combination with published photometry to estimate the total stellar mass of the host galaxy of PKS0347+05. In making this estimate we have used our lightly reddened OSP models to calculate the total stellar mass (OSP+YSP) within our spectroscopic aperture, and the r'-band photometry of \citet{ramos11} to calculate the mass of the stellar halo outside the spectroscopic aperture\footnote{We estimate that our spectroscopic aperture contains 26\% of the total stellar light at 6300\AA\, contained within the 30~kpc diameter photometric aperture used by \citet{ramos11}.}. In the latter case we assume that the light from the extended halo of the galaxy is emitted entirely by an unreddened OSP. Making this assumption we find total stellar masses in the range $9\times10^{11} < M_{host} < 1.3\times10^{12}$~M$_{\odot}$ --- similar to the  
$M_{host} = 1.3\times10^{12}$~M$_{\odot}$ obtained by comparing the 64~kpc diameter aperture K-band photometry of \citet{inskip10} with the spectral synthesis models of \citet{maraston09}. Alternatively, if we assume that the 
stellar population mix and reddening are the same in the extended halo
as in the nuclear aperture, we find that the total stellar masses
are up to a factor of two higher. Note that
all these mass estimates place the host galaxy of PKS0347+05 at the upper end of the mass range for both starburst and non-starburst radio galaxies \citep{seymour07,tadhunter11}. Indeed, PKS0347+05 falls significantly below the K-z relationship presented in \citet{inskip10} and, in terms of comparison with systems undergoing major mergers in the local Universe, it is more massive than the host galaxies of {\it any} of the 26 low redshift ULIRGs in the complete sample of \citet{zaurin10}.

\section{Discussion}

\subsection{Explaining the unusual properties of PKS0347+05}

PKS0347+05 shows an unusual combination of properties.
Although it is typical of powerful, FRII radio galaxies at intermediate redshifts in the sense that it is hosted by a giant elliptical galaxy that shows signs of a recent galaxy interaction
\citep[see][]{inskip10,ramos11}, it is unusual in showing a lack of evidence for AGN activity at both optical and mid-IR wavelengths, yet strong evidence for prodigious star formation activity. This rare combination of powerful FRII radio and strong recent ($t_{ysp} < 100$~Myr) star formation activity is generally associated with objects that also show 
quasar-like levels of nuclear emission line emission line activity ($L_{[OIII]} > 10^{35}$~W: Tadhunter et al. 2011).

The lack of evidence for powerful current AGN activity in the nuclear regions of PKS0347+05 is all the more suprising given that the system appears to be undergoing a major gas-rich merger that has triggered substantial star formation activity; under such circumstances one would expect there to be a plentiful supply of gas fuel to the nuclear regions of the host galaxy, as predicted by hydrodynamical simulations of
gas-rich mergers \citep{dimatteo05,dimatteo07,johansson09}. 

How do we explain the lack of powerful AGN activity in the nuclear regions of this system despite the evidence for powerful jet activity in the form of the FRII radio source? One possibility is that this is an extreme example of a FRII/WLRG in which the particular mode of accretion onto the black hole is such that it leads to powerful FRII radio activity without luminous emission line activity. For example, it has been suggested that WLRG are associated with fuelling via hot gas accretion, whereas SLRG are associated with cold gas accretion \citep[e.g.][]{hardcastle07,buttiglione10}. However, this explanation seems unlikely in the case of PKS0347+05 given that it is undergoing the final stages of a major gas-rich merger that is leading to prodigious star formation activity, and would be expected to be associated with a plentiful supply of cold gas fuel to the nuclear regions. Moreover, it is difficult to explain the jet power of such a luminous radio source purely in terms of Bondi accretion of the hot ISM, unless the central black hole is unusually massive for its K-band luminosity and/or the hot ISM in the central regions of the galaxy
is unusually dense \citep[see Figure 1 of][]{hardcastle07}. 

More plausibly, the jet and associated AGN activity in PKS0347+05 are indeed fuelled by cold gas accreted in the merger, but the fuel supply is intermittent. In particular, we suggest that rate of merger-driven accretion into the nuclear regions has recently diminshed substantially in this object on a timescale that is shorter than the light travel time time to the radio/lobes hot spots ($\sim5\times10^5$~yr). Therefore, at the time we are observing the source, the level of nuclear AGN/jet activity is relatively low --- leading to the WLRG classification --- but the information about the decreased AGN/jet activity has yet to reach the extended radio lobes/hot spots, which still reflect the old (higher) levels of nuclear activity. Indeed, many of the most recent hydrodynamical simulations of major, gas-rich galaxy mergers provide evidence that, despite the generally high rates of accretion into the nuclear regions in the final stages of such mergers, the gas flows are highly intermittent \citep[e.g.][]{johansson09}. Moreover, further evidence for intermittency of the fuel supply is provided by mid-IR spectroscopic observations of ULIRGs which, despite generally being considered to represent the final stages of gas rich mergers \citep[within $\sim$100~Myr of coalescence of the nuclei:][]{zaurin10}, show a considerable range in the level of nuclear AGN activity, with the AGN energetically dominant in $<$50\% of cases \citep{veilleux09}.
We also note that, based on the detailed analysis of their optical emission line spectra, it has been suggested that the AGN in some other WLRG have only recently switched off \citep{capetti11}. 

The other interesting feature of PKS0347+05 is that it represents a rare example of a radio-loud/radio-quiet double AGN. Although many double AGN systems are now known -- many detected on the basis of SDSS data -- radio-loud/radio-quiet double AGN are relatively rare, perhaps reflecting the fact that radio-loud AGN comprise only a small fraction of the overall AGN population, and are associated with giant elliptical galaxies at the upper end of the galaxy mass function. In this context we note that, of the 20 powerful radio galaxies from the $0.05 < z < 0.7$ 2Jy sample that show double nuclei or are tidally interacting with a companion galaxy \citep{ramos11}, as far as we are aware, only PKS0347+05 shows evidence for dual AGN activity (representing only 5\% of the pre-coalescence systems in the 2Jy sample). Previously reported examples of radio-loud/radio-quiet dual AGN include 3C294 \citep{stockton04}, 3C321 \citep{robinson00,evans08} and FIRSTJ164311.3+315618 \citep{brotherton99}. However, unlike PKS0347+05, the radio galaxy components in all of these latter systems show powerful nuclear emission line activity.

The simultaneous triggering of AGN in two galaxies undergoing an interaction can be naturally understood in terms of the radial gas flows driven by the tidal torques associated with such interactions. The radial gas flows are expected to take place simultaneously in both of the interacting galaxies, potentially triggering AGN activity. Given the large amounts of gas being funnelled into the near-nuclear regions of the interacting galaxies, the 
question then arises as to why dual AGN are not more common in interacting
systems. The answer is likely to relate to the intermittency of the fuel supply, as already discussed above. Certainly the presence of large quantities of gas in the central kpc is not by itself a sufficient condition for the triggering of an AGN: the gas must also shed a considerable amount of its angular
momentum to be accreted to the requisite sub-pc scales, and feedback effects associated with the central interaction-induced starburst component may either enhace of decrease the rate of accretion onto the central supermassive black hole.

Recent hydrodynamical simulations of the triggering of AGN in major spiral/elliptical mergers \citep{vanwassenhove11} demonstrate that dual AGN are likely to be observed in only a small percentage ($\sim$0.6 -- 16\% depending on the AGN luminosity thereshold) of pre-coalescence elliptical-spiral mergers for nuclear separations $>$10~kpc. This is consistent with the low rate of dual AGN activity among the pre-coalescence radio galaxies in the 2Jy sample\footnote{N.B. A caveat with comparing the simulations with the results for 2Jy and other radio galaxy samples is that the \citep{vanwassenhove11} simulations concern galaxies that have significantly lower stellar and black hole masses than typical radio galaxies; the extent to which the black hole mass affects the dual AGN statistics is not clear.}. 

\subsection{The lifetimes of intermediate redshift
FRII radio sources}

Assuming that our interpretation of the properties of PKS0347+05 is correct, by looking at the rate of occurrence
of similar objects in the radio source population, we can obtain an estimate of
the typical lifetimes of intermediate redshift FRII radio sources that is independent of the usual spectral aging and dynamical arguments. 

There are five timescales to consider: the lifetime of the FRII radio source ($\tau_{FRII}$), the lifetime of the associated strong line radio galaxy (SLRG)/quasar
phase of nuclear activity ($\tau_{SLRG}$), the lifetime of the weak line radio galaxy (WLRG) phase ($\tau_{WLRG}$), the light crossing time of the 
narrow line region ($\tau_{NLR}$),  and the light travel time from nucleus to the radio hotspots ($\tau_{HS}$). Given that the typical light crossing time of the near-nuclear narrow line region is $\tau_{NLR} \sim 3\times10^3$~yr (assuming $r_{NLR} = 1$~kpc), and typical powerful FRII radio sources have
hotspot light travel times 
$\tau_{HS} > 10^5$~yr, we deduce that
$\tau_{NLR} << \tau_{HS}$. 

Following the scenario we have proposed for PKS0347+05, we start by assuming that we observe WLRG optical spectra
in powerful FRII radio sources for the full period of time after the central AGN and jets have switched
off, and the NLR has faded, but before the information about this has reached the radio hotspots (i.e. over the time interval $\tau_{NLR} < t < \tau_{HS}$). In
this case $\tau_{\small FRII} = \tau_{SLRG} + \tau_{HS}$, $\tau_{HS} = \tau_{WLRG}$, and the fraction of WLRG detected in the population of powerful
FRII radio sources is then 
$f_{WLRG} = \tau_{HS}/\tau_{FRII}$. This
leads to the following estimate of the lifetime of a typical FRII
radio source:
$\tau_{FRII} = \tau_{HS}/f_{WLRG} = L_{rad}/(2 c f_{WLRG})$, where $L_{rad}$
is the typical linear size of a radio source just before it switches off. 

To determine the fraction of WLRG in the population of powerful radio galaxies
($f_{WLRG}$) we consider the sub-sample of radio sources in the $0.05 < z < 0.7$ 2Jy sample described in \citet{tadhunter98} and \citet{dicken09} with 5~GHz radio
powers $P_{5GHz} > 10^{26}$~W Hz$^{-1}$. This radio power cut ensures that
all the radio sources are genuinely powerful radio sources with FRII or
compact steep spectrum (CSS)/gigahertz peaked (GPS) radio morphologies\footnote{The only exception is PKS1648+05 (Her A), which has 
a radio morphology that is intermediate between FRI and FRII.}. 
We find that the  high radio power subset of the full 2Jy sample comprises a total of 29
objects, of which 26  are classified as SLRG, and only 3 
(PKS0347+05, PKS1648+05 and PKS2211-17) are classified as WLRG\footnote{Tadhunter et al. (1998) define WLRG as objects with [OIII] equivalent widths $EW_{OIII} < 10$~\AA. Although this definition encompasses objects in which no [OIII] emission is detected, there are no such objects in the high power 2Jy sample.}. Therefore
we deduce  $f_{WLRG} = 0.1$ and, assuming that $L_{rad} = 326$~kpc and $\tau_{HS}
= 5\times10^5$ are typical of radio galaxies of similar radio power
and  redshift
as  PKS0347+05 \citep[e.g.][]{allington84}, we obtain $\tau_{FRII} = 5\times10^6$~yr. This is consistent with spectral aging estimates of the lifetimes of powerful FRII radio sources \citep[e.g.][]{alexander87}.

One of the major assumptions we have made is that the sources will appear
as WLRG over the full period $\tau_{NLR} < t < \tau_{HS}$. However, if the
AGN has switched off completely over this period, and the AGN is the only source
of ionization for the [OIII] emission, then we would expect to observe negligible [OIII] flux in most WLRG. This is because the e-folding time of the intensity decline in the [OIII] emission is only $\sim$20~yr for a typical NLR electron
density of $n_e = 100$~cm$^{-3}$ \citep{binette87,capetti11}. In contrast,
we do detect weak but significant [OIII] emission from all  three WLRG in
the high power radio galaxy sample described above. A possible explanation for
this apparent contradiction is that there remains significant low-level
AGN activity that is capable of photoionizing the [OIII] emission line region in the post-SLRG phase. Alternatively, other ionizing sources (e.g. young stars
or shocks) may be sufficient to  produce the low-level [OIII] emission once
the AGN has faded. In this context it is notable that at least two of the
three WLRG in the high power radio source sample -- PKS0347+05 and PKS1648+05 -- fall in the ``composite'' region
of the diagnostic diagrams in which stellar photoionization may play a significant role.

Finally we emphasise that, if we are wrong in our assumption that {\it all} WLRG are objects
in which the AGN/jets have recently switched off, our estimated
FRII radio source lifetime represents a lower limit on $\tau_{FRII}$, because $f_{WLRG}$ --- representing
the fraction of genuine ``switch off'' WLRG sources in the above equations --- would then be lower
than we have assumed; if $\tau_{FRII}$ were substantially 
less that the $5\times10^6$~yr we have estimated, and all FRII radio
sources grow to a linear size of a few hundred kpc before their AGN switch off,
we would expect to detect a much higher fraction of WLRG in the high power radio source population than is observed.

\section{Conclusions}

We identify PKS0347+05 as a dual radio galaxy/Seyfert 1 AGN system in which 
both the AGN have been triggered as a consequence of the gas flows associated with a major galaxy merger. The fact that no powerful optical or mid-IR AGN activity is currently detected in the nucleus of the galaxy hosting the powerful FRII radio source can be explained in terms of a recent, rapid decline in the nuclear AGN activity within the last $10^6$ years which has yet to be reflected in the properties of the extended radio source. This, along with the overall rarity of dual AGN, highlights the intermittency of the AGN fuel supply, even in systems undergoing the final stages of major, gas-rich mergers.

\section*{Acknowledgments}

CRA and MR acknowledge financial support from STFC. CRA acknowledges the Spanish Ministry of Science and Innovation (MICINN) through project Consolider-Ingenio
2010 Program grant CSD2006-0070: First Science with the GTC. We thank the referee for useful comments that have helped to improve the manuscript.
Based on observations obtained at the Gemini Observatory, which is operated by the 
Association of Universities for Research in Astronomy, Inc., under a cooperative agreement 
with the NSF on behalf of the Gemini partnership: the National Science Foundation (United 
States), the Science and Technology Facilities Council (United Kingdom), the 
National Research Council (Canada), CONICYT (Chile), the Australian Research Council (Australia), 
Ministério da Ciência, Tecnologia e Inovação (Brazil) 
and Ministerio de Ciencia, Tecnología e Innovación Productiva (Argentina).
The William Herschel Telescope is operated on the
island of La Palma by the Isaac Newton Group in the Spanish
Observatorio del Roque de los Muchachos of the Instituto de
Astrofisica de Canarias. This research has
made use of the NASA/IPAC Extragalactic Database (NED) which is
operated by the Jet Propulsion Laboratory, California Institute of
Technology, under contract with the National Aeronautics and 
Space Administration.

\end{document}